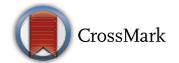

# A framework of blockchain-based secure and privacy-preserving E-government system

Noe Elisa[1] · Longzhi Yang[1] · Fei Chao[2] · Yi Cao[3]

**Abstract**
Electronic government (e-government) uses information and communication technologies to deliver public services to individuals and organisations effectively, efficiently and transparently. E-government is one of the most complex systems which needs to be distributed, secured and privacy-preserved, and the failure of these can be very costly both economically and socially. Most of the existing e-government systems such as websites and electronic identity management systems (eIDs) are centralized at duplicated servers and databases. A centralized management and validation system may suffer from a single point of failure and make the system a target to cyber attacks such as malware, denial of service attacks (DoS), and distributed denial of service attacks (DDoS). The blockchain technology enables the implementation of highly secure and privacy-preserving decentralized systems where transactions are not under the control of any third party organizations. Using the blockchain technology, exiting data and new data are stored in a sealed compartment of blocks (i.e., ledger) distributed across the network in a verifiable and immutable way. Information security and privacy are enhanced by the blockchain technology in which data are encrypted and distributed across the entire network. This paper proposes a framework of a decentralized e-government peer-to-peer (p2p) system using the blockchain technology, which can ensure both information security and privacy while simultaneously increasing the trust of the public sectors. In addition, a prototype of the proposed system is presented, with the support of a theoretical and qualitative analysis of the security and privacy implications of such system.

**Keywords** E-government systems · Blockchain technology · Cybersecurity · Privacy · Decentralized systems

## 1 Introduction

The wide availability of internet has motivated different countries around the world to exploit technologies as a means of communication and services exchange between citizens and other affiliates. E-government users enjoy the online services without leaving their comfortable homes, evading long queues in public offices whilst saving time and transportation costs, and at the same time the service providers can deliver services more effectively and efficiently. Generally speaking, government networks can communicate to each other better than business networks, because most of them are connected for transferring information to the public without competition. In the future, the number of devices using e-government services will increase dramatically due to the fast evolution of smart homes, internet of things (IOT), smart cities, and other interconnected networks [1, 2]. According to the United Nation e-government survey, 2014 [3], almost every government around the world is currently providing its citizens and other stakeholders e-services via websites and mobile applications.

E-government systems collect, store and process a significant amount of confidential information about citizens, employees, customers, products, researches, financial

✉ Longzhi Yang
longzhi.yang@northumbria.ac.uk

Noe Elisa
noe.nnko@northumbria.ac.uk

[1] Department of Computer and Information Sciences, Northumbria University, Newcastle upon Tyne NE1 8ST, UK

[2] Cognitive Science Department, Xiamen University, Xiamen, People's Republic of China

[3] Technology Department of Nanjing Customs District, Nanjing, People's Republic of China

status amongst others, using electronic computers. The compromise of such information usually leads to the loss of users' trust and confidence, opportunities, and financial advantages, etc [4]. It has been found that more than 80% of e-government web sites around the globe were vulnerable to cross-site scripting (XSS) and structured query (SQL) injection due to the lack of proper authentication mechanism applied to input data from users [5]. Recently, many nations around the world have experienced a great threat from denial of service attacks (DoS) and malware targeting their networks [6]. For instance, the U.S. government suffered one of the largest e-government attacks in 2015, causing the leak of over 4 million government employees' confidential information, including security clearance information, social security numbers, identities and passwords [7]. According to the report in [8], in 2016, the Tanzania government was hit by cyber-terrorists, technology spies, hackers and digital fraudsters leading a loss around 85 millions US dollars. In addition, more than 1500 user accounts in Singapore were hacked in the government platform in 2014, where hackers gained access to creating new businesses and apply for work permits [7].

It is therefore of significant importance to ensure the security, privacy, confidentiality, integrity and availability of e-government systems. The existing e-government systems, such as e-government websites and eIDs management systems [9, 10], are centralized where one or duplicated central servers and databases store and provide information to users. The centralized management and validation system is likely to suffer from a single point of failure and makes the system a target to cyber attacks such as DDoS, DoS and malware. Any e-government system will remain vulnerable to privacy and security breaches if better security technology and countermeasures are not developed and available to combat these threats in the future.

The blockchain technology has appeared to be one good solution to provide a secure decentralized environment for information exchange [11, 12]. Although it was originally introduced for exchanging digital currency as its underlying technology, it has found security and privacy applications in many other areas, such as Internet of Things (IoTs) [13], smart home [14], smart city [1], educational systems [15], and healthcare [16]. While governments around the world have not fully adopted the blockchain technology in the public sectors, many countries have initiated blockchain projects to explore the potential of blockchain technology in offering public services to individuals [17]. Each of these projects usually focuses on a particular service, for example e-residency, e-health, land registry etc; they are still in their early stages and no common framework has been proposed for blockchain technology integration within the e-government systems [17]. In addtion, each of these countries is developing their own blockchain framework. Different blockchain systems from different countries lead to difficulties to communicate out of their network for international information exchange.

This paper presents a framework and prototype of blockchain-based secure and privacy-preserving e-government system, which can be adopted by any government for the purpose of ensuring both security and privacy while simultaneously increasing trust in the public sector. The system is made of a peer to peer network of e-government devices (nodes) and user's devices. Briefly, any new e-government device or individual device joining the system will be reviewed by the existing peers of the network and one of the peer is elected to set up a network node and blockchain address of a new device. When a new user attempts to register with the system through a device or one of the government departments, the user is assigned with a user ID and a blockchain wallet for collecting and storing his/her transaction. By doing so, e-government users can submit and access their records from anywhere and everywhere, using their IDs and blockchain addresses. This paper has also analysed and evaluated the security and privacy implications of the proposed system in public sectors using theoretical and qualitative analysis.

The reminder of the paper is structured as follows. Section 2 briefs the background technologies on the blockchain and e-government systems. Section 3 introduces the proposed e-government framework and prototype. Section 4 discusses the security and privacy implications of the proposed system, and Sect. 5 concludes the paper and outlines the future research directions.

## 2 Background

The background technologies on the blockchain and e-government systems are reviewed in this section.

### 2.1 Blockchain technology

Blockchain is a peer to peer (p2p) distributed database (i.e., ledger) maintaining a list of continuously growing records called blocks that are linked and secured usually using public key cryptography [11]. By the blockchain technology, new information is added to a block and becomes available to all nodes in a distributed network, rather than adding to the centralized database in the traditional centralized system. Each block in a blockchain is identified by a hash value generated by usually using the secure hash cryptographic algorithm-256bits (SHA256) [11]. The hash value of a current block header (parent) is linked and stored in the next block (child) as depicted in Fig. 1 [18]; therefore, if there is an alteration in any block's content, its hash will also be changed accordingly and this change will be propagated throughout the network to invalidate that block [11]. Based on this mechanism, the blockchain technology does not require an intermediary or trusted third party as it is decentralized and distributed. The blockchain participants have private keys assigned to them to digitally sign and validate the transactions they make.

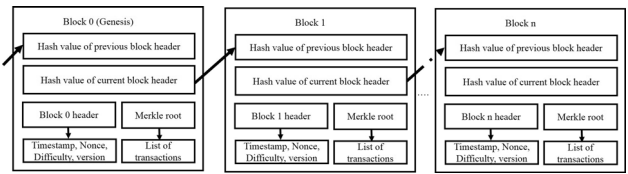

**Fig. 1** An example ledger with details of blocks

As shown in Fig. 1, a block is composed of a header containing the meta data, and a long list of transactions performed in that block. The block header typically contains the timestamp, nonce, version and proof of difficulty. The timestamp indicates the time that the block was created; the nonce is a random number generated by the consensus algorithm to compute the hash value of a block; the version indicates the version number of the blockchain; and the proof of difficulty is a generated hash value which must be less than the current target hash value.

The first block, known as a genesis block, is hardcoded by embedding some random data into the blockchain application [11]. Although each block has only one parent and one child, a valid block may have two or more children temporarily created when two or more nodes (network peers) are added to a block at the same time leading to two or multiple branches from the same parent [18]. This situation is usually called 'fork' and is eliminated by taking the chain whichever becomes longer than the others as a valid blockchain, and making all other shorter ones invalid (orphan), with a two-branch situation demonstrated in Fig. 2. It is possible that the formed branches have the same length; in this situation, the process of adding new blocks continue for all the to-be-validated chains until one branch becomes longer than the others thus valid.

Within a block, all the transactions are linked together using a merkle tree [11]. A merkle tree is an upside down binary tree used by the blockchain technology to summarise all the transaction in a block. To construct a merkle tree, a pair of transactions are hashed recursively until they form only one root node at the top of the tree termed as the merkle root [11], as shown in Fig. 3. More precisely, a merkle root is the hash of all the transactions that make up a block in a blockchain network. Any tiny modification of the data will change the merkle root hash leading to an invalid record. The most common cryptographic hash algorithm used to construct a merkle tree is the secure hash algorithm 256-bits (SHA256). If there is an odd number of transactions, the last transaction hash is duplicated to create an even number of transactions thus ending up with a balanced tree.

Nodes in the blockchain network run a consensus algorithm to validate transactions. There are several consensus algorithms (protocols) available for the blockchain technology, such as proof of work (PoW), proof of stake (PoS), delegated proof of stake (DPoS), and proof of difficulty (PoD) [18, 19]. For instance, Bitcoin employs the PoW while Ethereum and Bitshare implement PoS and DPoS, respectively [19]. In PoW, miner nodes which want to add (mine) a new block to the blockchain network must first solve a difficult mathematical puzzle which requires great computational power. The miner who solves the puzzle adds a new block and gets rewards in bitcoins [18].

Unlike the PoW, with the PoS, a node which creates a new block is chosen deterministically depending on its stake (wealth) [19]. PoS saves energy that is required in PoW to solve the mathematical puzzle, and only the wealth of a node (validator) is required to validate the new transactions and blocks. The DPoS attempts to solve the consensus problem using delegates [19]. DPoS uses a real time voting and reputation system to create a panel of limited trusted delegates who will witness and validate the blocks. The witnesses have the rights to create blocks and add them to the blockchain network, in addition to prohibit malicious nodes from participating in adding blocks. Principally, in PoS and DPoS, stakeholders of the network shares are not expected to deliberately make bad decisions for the network.

Blockchain network can be permissioned or permissionless [20]. A permissionless network or public blockchain allows any user to create a personal address, join the network, and participate in consensus, whilst a permissioned or private network only allows a number of restricted nodes to join.

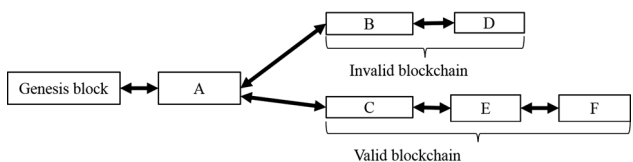

**Fig. 2** Blockchain validation for the fork situation

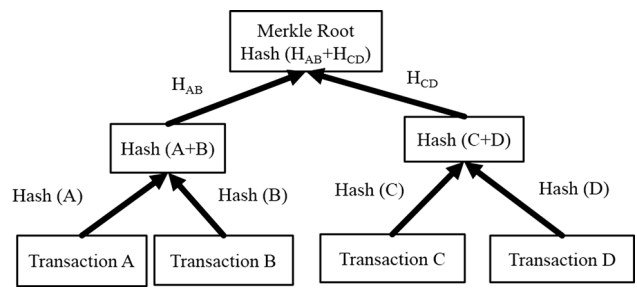

**Fig. 3** An illustration of the merkle tree

## 2.2 E-government systems

E-government services can typically be categorized into 3 groups: government to government (G2G), government to citizen (G2C), and government to business (G2B) [21]. G2G offers an online interaction between government departments, authorities, organisations and other governments to disseminate information among themselves using the internet. G2C and G2B allow citizens and businesses to interact with the government to get online services such as filing property tax, filing income tax, extending/renewal of visa, passport and licenses, online voting, e-procurement application, etc.

Every e-government department ensures that only authorized users can get access to individual's confidential information. In additional to security and privacy, another important factor to consider during e-government implementation is to build a trustable system that users can rely on [22]. Security and privacy assurance in e-government systems plays a critical role in increasing trustness between different departments within the government as it guarantees confidentiality, integrity, and privileged access of sensitive information. Typical attacks that e-government system faces include packet sniffers, probes, malware, DDoS, phishing etc [23, 24]. Due to cyber warfare, there are other new motivations for attacks such as political differences, extortion, cyber terrorism, and even contests for the supremacy which can occur within a nation or between different nations [6].

Different non-technical e-government security maturity models have been proposed for guiding and benchmarking the security implementation of the e-government system. For instance, a comprehensive e-government information security maturity model for guiding the inclusion of security in e-government systems was reported in the work of [21]. This model focuses only on the organisation's security mechanisms setting, security evaluation, security policy setting and information security awareness to users, but it lacks guidance for built-in security that can ensure e-service security and privacy.

Commonly, the security and privacy of e-government systems are provided using firewalls, intrusion detections systems (IDSs), public key infrastructure (PKI) and anti-virus mechanisms [2]. Various artificial intelligence techniques have been employed in the implementation of IDSs that can also be used by e-government systems [25–30]. In addition to these common solutions, one hardware solution to e-government security issues is the use of eID system [31] for identification, authentication, confidentiality and integrity of users' information. An eID system uses a smart card which includes a chip to store cardholder's personal data including date of birth, civil status, parenthood, current and past addresses, etc, in additional to certificate for authentication and digital signature [32]. An eID system provides a means for distinguishing between different citizens and businesses uniquely in order to access electronic services. eID cards are offered by migration boards and national identity register in most countries. The same eID can be used in multiple sectors (e.g. taxation, social security, education, telephony services, banking services) while fulfilling different roles (e.g. a civil servant or a businessman) depending on the context. The existing eID management systems are not interoperable and lacks security as a result of having single centralized system for keeping records [32].

Another technical solution for security and privacy issues in e-government is an authentication framework proposed in [33]. This framework utilises different registration and authentication procedures based on a single central citizen portal interfaced with ministerial departments. The framework consists of two parts namely Identity Provider (IdP) and the Service Provider (SP). Users have to register with IdP at a central portal of an e-government system to obtain unique identities that will be used to access services from the service providers. Whenever a user wants to access a service, SP must authenticate his/her identity with the IdP at the central portal to obtain a Single Sign On (SSO) password that can be used to access different SPs.

## 3 Blockchain-based E-government system

The proposed blockchain-based e-government framework is illustrated in Fig. 4. In this figure, a double direction arrow G2G shows an interaction taking place between government departments and organizations, allowing a peer to peer (p2p) exchange (broadcast) and validation of data being submitted by individuals. A G2C double direction arrow represents information sharing between citizens and the government such as filling forms for tax, marriage certificates, business permits, birth certificates, visa or passport. A G2B double direction arrow expresses information exchange such as electronic procurement, tax and insurance clearance forms and electronic auctions between public and business organizations (enterprises) as the main source of economic growth.

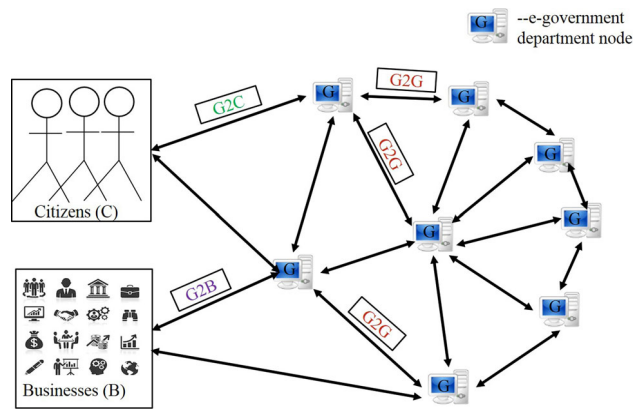

**Fig. 4** The proposed blockchain-based e-government network

The joining of any new e-government (G) node or user node (C or B) to the blockchain network is reviewed by peers in the network, and e-government tokens are assigned for setting up its network node, leading to a permissioned (private) blockchain. The number of e-government tokens is equivalent to the total number of records stored by a node in the blockchain network. Each user has a dedicated e-government blockchain wallet to collect his/her tokens. Once a record is submitted, the record will be transferred in terms of stake to his/her blockchain address. Using a DPoS protocol, any node will be able to register with the network as a delegate as described in Sect. 3.2. In order to add a new block to the blockchain, the departments that jointly make up the e-government system vote for a delegate who will validate the transactions and seal the block. This approach is appropriate for security requirement, since random nodes cannot join the network, generate new tokens and set up a network node unless the rest e-government nodes approve it. The permissioned blockchain system ensures that the stored records are trustworthy, auditable and transparent.

### 3.1 Network nodes

There are two types of nodes in blockchain terminology, which are full nodes and lightweight nodes [18]. A full node downloads a full copy of the blockchain when it joins the blockchain network, which allows it to fully validate transactions and blocks. A lightweight node does not download a complete copy of the blockchain when it joins the network, but it downloads only the block headers to validate the authenticity of transactions. To be able to transfer their transactions to the network and receive notifications when transactions affect their blockchain wallets, lightweight nodes usually refer to a copy of a trusted full node of the blockchain.

In the proposed system, e-government department nodes serve as full nodes while user's devices (C and B) serve as lightweight nodes, although any business node is allowed to download a complete copy of the blockchain. The p2p network connectivity in the proposed network can be provided by using wireless broadband, thanks to the fact that many countries across the world are trying to incorporate a city-wide wireless broadband network across the city using the Wi-Fi technology [34].

### 3.2 Delegates and witnesses voting

Peers need to agree on a state of the block's transactions and the process of sealing the blocks into the blockchain, in order for the network to remain functional. The DPoS is adopted here as the consensus algorithm due to its computational efficiency in adding transactions and sealing a block [19]. Briefly, the DPoS can be viewed as a representative democracy where the e-government nodes utilize their stake (records) in order to elect delegates (other nodes) to join the network. Delegates are responsible for securing the network and voting for witnesses among them who will validate transactions and seal the blocks.

To elect the witnesses, votes are weighed according to the size of the tokens of each delegate or voting node. A delegate does not need to have a large token to get selected as a witness, but votes from delegates with large tokens can result in delegates with relatively small tokens being elected as witnesses. E-government nodes will store its individual's records, allowing them to have higher tokens to participate and vote for witnesses which increases the security of the proposed system. In DPoS, nodes are allowed to delegate their voting power to other nodes, whom they trust to vote for witnesses on their behalf. Any delegate which appears to misbehave will be removed from the set of participating delegates and its votes is assigned to a new node joining the network or any existing node in order to ensure security and trust within the network.

The choice of the underlying blockchain technology for the prototype implementation mainly relies on the availability and efficiency of the DPoS implementation. In the meantime, the computational energy required for validating transactions must be affordable while the security of the established network must be guaranteed. For instance, Ethereum platform implements DPoS and smart contract protocol to simulate real contracts such as tax and insurance contracts, employment contracts, and land registry [35]. It provides an important alternative in e-government to store citizens' sensitive records, due to its ability to facilitate contract negotiation, simplify contract terms, implement contract execution, and verify contract fulfillment state.

### 3.3 New node creation

The process of registering a new node to the proposed e-government network is summarised in "Algorithm 1". Any e-government department can join the blockchain network by setting up a full network node while other users can only set up lightweight nodes. Once a new node join the network, a functional node will generate its blockchain wallet and address containing public and private keys as shown in lines 7 and 8 in "Algorithm 1". The private key is used by each node to sign and validate transactions therefore it must be kept safely (line 9). After generating the address, a node will contact delegates in the blockchain network to send its registration request, and one of the delegates will verify its registration and transfer some e-government tokens (registration record) in its blockchain address.

Thereafter, a new node is added to the network, and its registration is broadcasted to the network peers by the assigned delegate (line 12–14), allowing other network peers to receive its wallet information for sending transactions in the next cycle. Additionally, a new node receives instructions for a network node setup so that it can be elected as a delegate to validate transactions in the next cycle. Subsequently, the new node sets up the network node according to the instruction provided. In particular, the instruction consists of the size of initial token, the blockchain address of a node, and the public and private keys for signing and validating transactions before adding a block.

The process of adding a new node is completed when a network node is successfully set up and broadcasted to the network by a delegate. Information security is enhanced by using encrypted data distributed across the network. Therefore, even if a malicious node is registered as a department node, it cannot alter the data as every participating peer in the network is able to detect the alteration and invalidate the change.

**Algorithm 1** Adding a new node to the e-government network

**input:** Node registration request,
 Nodes $N$ in the current network
**output:** A newly created node $m$ in the e-government network
1: **if** (the request is from government) **then**
2:    create a full node $m$;
3: **else**
4:    create a lightweight node $m$;
5: **end if**
6: $(K_{pub}, K_{pr}) \leftarrow generateKeys()$
7: $Addr \leftarrow createBlockchainAddress() + (K_{pub}, K_{pr})$;
8: $Walt \leftarrow createBlockchainWallet() + (K_{pub}, K_{pr})$;
9: $safelyStorePrivateKey()$;
10: $Addr \leftarrow Addr+$tokens;
11: $\beta \leftarrow selectDelegate(N)$;
12: **for each** $n \in \{N - \beta\}$ **do**
13:    $distributeRegistration(n, m)$;
14: **end for**
15: $m \leftarrow verifiedNewNode()$;

### 3.4 User registration

Users make their registrations using their network devices or physically visiting one of the government departments, with the process summarised in "Algorithm 2". As described in lines 2 and 3 in this algorithm, an ID for a user is issued and a new blockchain address is generated for the user containing public and private keys, allowing the identification of the owner. A Blockchain wallet for this new user is created and broadcasted (lines 5–8) so that each node can store it in its blockchain address. The created blockchain wallet is then used to send and receive transactions related to the user's account. User IDs and private keys will be stored safely in the wallet file or the database of the user's device. Users can conveniently view their records and the new transaction available in their blockchain addresses through the wallet interface.

**Algorithm 2** Registering a new user

**input:** User registration request
 Nodes $N$ in the current network
**output:** A newly registered user $u$
1: $(K_{pub}, K_{pr}) = generateKeys()$
2: $uID \leftarrow createUserID()$;
3: $Addr \leftarrow createBlockchainAddress() + (K_{pub}, K_{pr})$;
4: $(uID, K_{pr}) \leftarrow Safelystore(uID, K_{pr})$;
5: $Walt \leftarrow createBlockchainWallet() + (K_{pub}, K_{pr})$;
6: **for each** $n \in N$ **do**
7:    $distributeWallet(n, Walt)$;
8: **end for**
9: $u \leftarrow verifiedNewUser()$;

When a user submits a record to a delegate, the transaction is authenticated and initialized. From this, the block is updated to a new version which is broadcasted across the network for validation and then transferred to his/her blockchain address in all network peers. The transferred record is stored in the blockchain address of the user with the following data content: (1) the ID of the user, (2) the record value such as property registration, and (3) the record identification such as tax registration number. Each data instance in a blockchain represents an asset.

When a third party organization (e.g., business) requests to access a user's information for any official issues, the user needs to provide his/her blockchain address for verification. The organisation can then use the blockchain web API to access the blockchain data stored in the user's address. All e-government users are required to backup their private keys and keep them safe. If any user lost his/her private key, (s)he will be required to create a new blockchain address and request one of the e-government department node to transfer his/her information from the old blockchain address to a newly created blockchain address.

The user's device and identity will be validated and authenticated, when a registered user wants to access the network. This helps to minimise human errors which have always been considered as a main contribution of failure and a weak link to access information stored in information systems [36]. As a result, government information will flow securely and seemingly to the right individuals at the right time and right place. Typical human errors in cyber security include sending sensitive data to the wrong recipient and unintentionally exposing login credentials such as username and passwords. According to [37], human error remain as one of the main causes for cyber security breach in the public and private organisations.

### 3.5 New block generation

Every block is created by one active witness who is selected at random by the majority of peers from the list of active delegates. If a witness misses a block, another witness is tasked to create and validate the block to join the blockchain network. In DPoS, a fixed period of time is set for block creation, often five seconds [19]. "Algorithm 3" highlights the fundamental steps involved in the process of generating and adding a block to the blockchain using the DPoS consensus algorithm.

---

**Algorithm 3** Creating and adding a new block to the blockchain

**input:** A set of $N$ nodes in the current network,
  A blockchain $B$ including blocks from $b_0$ to $b_n$,
  Consensus time $T_c$ required to create a new block
**output:** A newly created block $b_{n+1}$
1: Initialize an empty set of transaction $R = \{\}$;
2: $\beta \leftarrow WitnessElection(N)$;
3: **while** transaction_time $< T_c$ **do**
4:   **for each** $n \in \{N - \beta\}$ **do**
5:     $R \leftarrow R + GetTransactionsfromNode(n)$;
6:   **end for**
7: **end while**
8: $b_{n+1} \leftarrow createBlock(b_n, R)$;
9: **for each** $n \in \{N - \beta\}$ **do**
10:   $signBlock(b_{n+1}, n)$;
11: **end for**
12: $B' \leftarrow B + b_{n+1}$;
13: **for each** $n \in N$ **do**
14:   $distributeBlockchain(B', n)$;
15: **end for**

---

A block is added in a regular interval of time, $T_c$. Within this interval, the block undergoes the following phases of activities. First, an empty set of transactions $R$ is initialized; and one witness from a group of delegates is elected to create and validate the transactions for the blockchain. Secondly, all the transactions are sent to the elected witness. This process continues until the witness stops accepting any new transactions for the block. Thirdly, the witness assembles the new block and distributes it to the network delegates for review and verification. This allows those nodes that elected the witness to digitally sign the block to prove its correctness. The signed block is returned to the witness and added to its local blockchain while simultaneously distributing the new block to the network. The witness cannot mine its own transaction and hence in lines 4 and 9, $\beta$ (witness) is excluded from the set of nodes $N$. At the end of the algorithm, the blockchain is distributed to all government nodes in the network.

## 4 Security and privacy analysis

Every e-government system must guarantee the confidentiality, integrity and availability of the services. Confidentiality is achieved when information is not disclosed to unauthorised users; integrity is achieved by protecting information from any form of modification, whilst availability means information is available when needed and is free from DoS or DDoS or other similar service disruption. This section provides a theoretical qualitative analysis on the security and privacy performance of the blockchain-based e-government system.

The records stored in the proposed system are secured through the public key cryptography that protects against adversarial attempts to alteration and/or unauthorised access, whilst network users are assigned with private keys for validating and signing transactions. Encryption and digital signature are used in the network to ensure security, privacy and access control to the stored records. Moreover, most of the blockchain consensus algorithms (in this case DPoS) require an attacker to control at least 51% of the network peers in order to alter a record [20], which is generally impossible to achieve. More precisely, in order to change any block in the blockchain, an attacker has to modify every copy of that block in the network and then convince most of the nodes that the new block is the valid one. Also, to increase the privacy of the data stored in the proposed network, all user's blocks are hashed and an incomprehensible hashes of the transactions are stored in the blockchain.

The proposed system is a decentralized p2p system where the user's data are stored in different nodes which

Table 1 Security services and common measures

| Security service | Countermeasure (s) |
|---|---|
| Authentication | Blockchain address and digital signature |
| Access control | Digital Signature and encryption |
| Confidentiality | Encryption |
| Integrity | Encryption and digital signature |
| Non-repudiation | Encryption and digital signature |
| Availability | Distributed/decentralised |
| Trust | Decentralized, encryption and digital signature |

Table 2 Comparison between RSA and ECC key lengths in bits

| RSA key length | ECC key length | Approx. ratio (RSA:ECC) |
|---|---|---|
| 1024 | 160 | 6:1 |
| 2048 | 224 | 9:1 |
| 3072 | 256 | 12:1 |
| 7680 | 348 | 20:1 |
| 15360 | 512 | 30:1 |

The security services and corresponding common measures provided by the proposed framework is summarised in Table 1, which ensure adequate privacy and security of the transactions. For computational efficiency, user's devices will run lightweight client to store the transactions rather than the complete copy of the blockchain which is expensive in terms of storage. E-government devices are expected to be computationally powerful with enough storage capacity to store and process user's records efficiently. The network is able to offer the performances provided by the blockchain technology and DPoS consensus protocol such as scalability, speed, interoperability, and transparency, and it can handle a large number of transactions.

The elliptic curve cryptography (ECC) approach is adopted for implementing encryption and digital signature in the proposed framework, which is a common practice for most of the existing blockchain technologies such as Bitcoin and Ethereum. Note that the ECC and the RSA (Rivest-Shamir-Adleman) offers a similar level of security but ECC consumes far less number of bits [38]. For instance, a 256-bit key in ECC offers the same level of security as that provided by the RSA using a 3072-bit key. Shorter key usually means low CPU consumption, low memory usage and fast key generation. These advantages are also beneficial to the proposed framework in facilitating fast creation of the transactions and sealing the blocks. A summary of key length study between the RSA and ECC is provided in Table 2 [38]. 256-bit ECC keys are widely in guarantees the availability of the system by avoiding any single point of failure. Using DPoS, it is difficult for any adversary to launch DDoS or DoS attacks against the system as registration is required for a node to start sharing information with the rest of the network peers. Any transactions received from the network node are validated by the witnesses making it difficult for malicious nodes to initiate malicious connections.

Table 3 Features of blockchain-based e-government system

| Feature | Justification |
|---|---|
| Reduced human errors | User devices and identities are authenticated in advance before gaining access to the network |
| Increased public trust | Individuals have direct control of their information and all network participants are authenticated |
| Greater scalability | The system can easily scale up as it allows new devices and users to be added to the network automatically following the consensus mechanism |
| Improved reliability | Data are stored in multiple servers/locations. The consensus protocol ensures that data can only be altered when all participants agree so |
| Increased resiliency | Single point of failures are avoided and the system hence resilient to malware, DoS and DDoS attacks |
| Improved auditability | It is easy to trace back the history of all transactions since they remain unchanged in the network |
| Greater verifiability | All new transactions are validated by all participating peers in the network before being added to the blockchain |
| Information ownership | Individuals are responsible to authorise who will access their information |
| Improved access to information | Information is stored at multiple locations which enhances easy and speed access |
| Increased data quality | All transactions and records stored in the system are validated in advance making the stored information authentic with the required quality |
| Greater transparency | All nodes in the network share the same copy of the blockchain and new transactions are added based on the consensus mechanism |
| Reduced operational costs | There is no third party organisation needed to process transactions |
| Improved efficiency and speed | Anyone in the network are able to access to all records subject to the accessibility privilege and new records are distributed to all participating nodes |

blockchain technology as they can provide the required level of security for the majority of applications.

Apart from security and privacy preserving, the blockchain-based e-government system also provides a number of other benefits as summarised in Table 3. These features make blockchain technology as a perspective trend in the implementation of an e-government system, which is able to provide a convenient, safe and fault tolerant communication channel between the public sectors and citizens. The indirect benefits brought by the blockchain technologies, such as bureaucracy reduction, paper usage exclusion, transaction costs reduction, and corruption control, can change the governace ecosystem with higher degree of trust from citizens.

# 5 Conclusion and future work

This paper proposed an e-government framework that can enforce security and privacy in the public sectors by employing the blockchain technology. The theoretical and qualitative analysis on security and privacy of the framework shows that, cryptography, immutability and the decentralized management and control offered by the blockchain technology can provide the required security and privacy in e-government systems. The proposed system also has the potential of solving the interoperability issues between governance departments which is one of the limitations of existing e-government systems. As this work is limited at the framework and theoretical discussion level, the active future work is to implement such framework and then further explore its full potential in a real-world environment. Note that, the blockchain technology such as Ethereum is still at its early stages of development and therefore another piece of future work would be the application of the appropriate version of blockchain technology in public sectors to meet and increase the security and privacy of individual's data.

**Acknowledgements** This work has been supported by the Commonwealth Scholarship Commission (CSC-TZCS-2017-717) and University of Northumbria at Newcastle, UK.

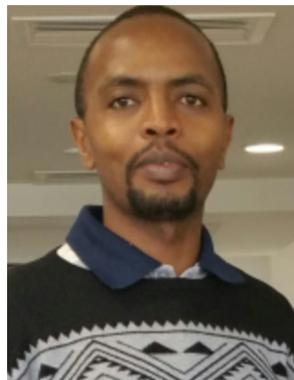

**Noe Elisa** is currently a PhD student in the Department of Computer and Information Sciences at Northumbria University, UK, sponsored by the Commonwealth Scholarship (No. TZCS-2017-717). He received his M.Tech degree in Computer Networks and Information Security from Jawaharlal Nehru Technological University, India, in 2014. Prior to that, he received a B.SC. degree in Telecommunication Engineering from the University of Dar es salaam, Tanzania in 2010. He has more than seven years of teaching experience with the University of Dodoma, Tanzania, where he worked as an assistant lecturer. His research interests include information security, privacy assurance, e-government systems and smart cities, blockchain technology, computational intelligence, machine learning, human computer interaction and cloud computing.

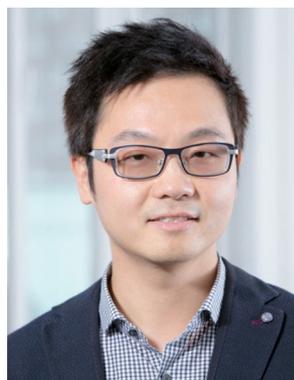

**Longzhi Yang** (MBCS, SMIEEE, FHEA), is an Associate Professor (Reader) and Director of Learning and Teaching in the Department of Computer and Information Sciences at Northumbria University. His research interests include artificial intelligence, machine learning and the applications of such techniques in real-world uncertain environments such as robotics, cyber-security, computer visions, and smart manufacturing. He is the chair of IEEE Special Interest Group of Big Data for Cyber Security and Privacy. He and his principally supervised students were the recipients of three best paper awards and several IEEE Computational Intelligence Society outstanding student paper travel grants.

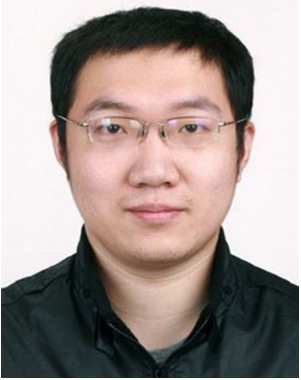

**Fei Chao** received the BSc degree in mechanical engineering from Fuzhou University, Fuzhou, China, in 2004, the MSc (Hons.) degree in computer science from the University of Wales, Aberystwyth, U.K., in 2005, and the PhD degree in robotics from Aberystwyth University, Wales, U.K., in 2009. He was a Research Associate with Aberystwyth University from 2009 to 2010, under the supervision of Prof. M. H. Lee. He is currently an Associate Professor with the Cognitive Science Department, Xiamen University, Xiamen, China. He has authored over 30 peer-reviewed journal and conference papers. His current research interests include developmental robotics, machine learning, and optimization algorithms. He is also the Vice Chair of the IEEE Computer Intelligence Society Xiamen Chapter. He is a member of the China Computer Federation.

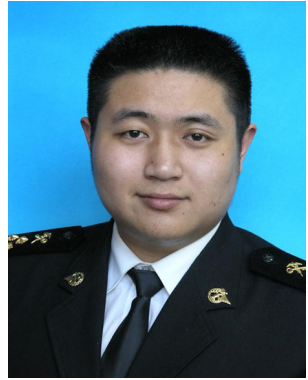

**Yi Cao** received his bachelor's degree in Computer Science and Technology from Nanjing University of Science and Technology. He has been involved in many software development projects, specialising in application system architecture design. His research interests include system architecture, software engineering, and cybersecurity.